# Introduction to Analytical Software Engineering Design Paradigm

1st Tarik HOUICHIME  
*ENSIAS, SPM Laboratory*  
*Mohammed V University In Rabat*  
Rabat, Morocco  
tarik_houichime@um5.ac.ma

2nd Younes El Amrani  
*ENSIAS, SPM Laboratory*  
*Mohammed V University In Rabat*  
Rabat, Morocco  
y.elamrani@um5r.ac.ma

*Abstract*—As modern software systems expand in scale and complexity, the challenges associated with their modeling and formulation become increasingly intricate. Traditional approaches often fall short in effectively addressing these complexities, particularly in tasks such as design pattern detection for maintenance and assessment, as well as code refactoring for optimization and long-term sustainability. This growing inadequacy underscores the need for a paradigm shift in how these challenges are approached and resolved. This paper presents and explores Analytical Software Engineering Design Paradigm (ASE), a novel and pioneer design paradigm designed to balance abstraction, tool accessibility, compatibility, and scalability, enabling the effective modeling and resolution of complex software engineering challenges. The ASE design paradigm is then evaluated through an analysis of the results obtained from two frameworks—Behavio-Structural Sequences (BSS) and Optimized Design Refactoring (ODR)—both developed in accordance with ASE principles. BSS provides a compact, language-agnostic representation of codebases, facilitating precise design pattern detection. Meanwhile, ODR unifies artifact and solution representations, optimizing code refactoring through heuristic algorithms while eliminating iterative computational overhead. By offering a structured approach to software design challenges, ASE establishes a foundation for future research in encoding and analyzing complex software metrics.

## I. Introduction

Software engineering is fundamentally a problem-solving discipline, yet the increasing complexity of modern software systems challenges the efficiency and adaptability of existing methodologies [1]. Traditional approaches often focus on refining tools to accommodate more intricate problems, leading to ever-growing computational overhead and diminishing returns. This raises a fundamental question: Should we continuously refine tools to fit complex problems, or should we rethink how problems themselves are formulated? For instance, Devanbu et al. [2] highlight that refining existing tools to manage ever-growing codebases often leads to significant computational overhead with diminishing returns. Similarly, while Component-Based Software Engineering (CBSE) has improved modularity and reuse [3], it remains limited by its dependency on rigid, tool-specific implementations. Model-Driven Engineering (MDE) offers advantages in abstracting system functionality [4], yet its reliance on domain-specific configurations frequently hampers its applicability across diverse problem domains.

Analytical Software Engineering (ASE) Design Paradigm, introduces a paradigm shift by emphasizing problem abstraction and compatibility over tool adaptation. Instead of tailoring tools to fit increasingly complex problems, ASE advocates for restructuring problems into representations that align with robust, well-established computational techniques. This enables the seamless integration of machine learning models, optimization heuristics, and mathematical formulations without the need for specialized tool adaptation.

At its core, ASE is driven by abstraction, generality, and scalability. By carefully encoding software engineering problems into generic representations, ASE preserves essential problem characteristics while reducing unnecessary complexity (See Figure 1). This design paradigm enables a systematic, computation-efficient approach to software design, analysis, and optimization, fostering a bridge between theoretical problem modeling and practical, scalable solutions. This paper formalizes the ASE paradigm, detailing its principles, core methodologies, and the fundamental shift it introduces to software engineering research and practice.

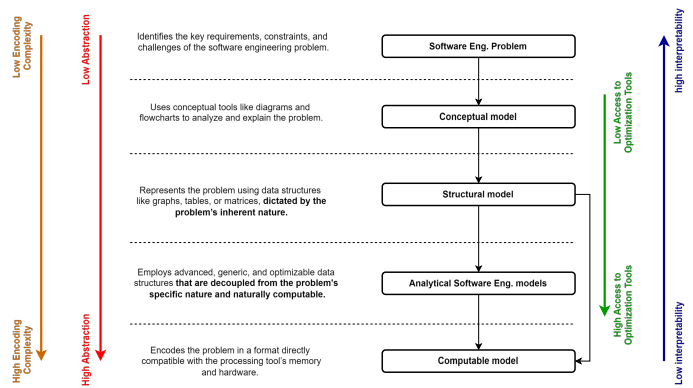

Fig. 1: Positioning of ASE paradigm models within the general framework for problem abstraction and representation in software engineering, considering abstraction levels, encoding complexity, interpretability, and tool accessibility.

## II. Definition and Principles of Analytical Software Engineering Design Paradigm

### A. Definition

We define the design paradigm of ASE, as the structured process of *"encoding and representing a software engineering problem in a manner that enables the problem to be processed and analyzed using a broad range of mathematical, analytical, and algorithmic tools, without requiring significant adaptation of those tools to fit the problem"*.

### B. Principles of ASE

The ASE paradigm is based on several core principles that guide its approach to solving software engineering problems:

- **Abstraction and compatibility:** The process involves distilling complex problems into abstract representations while maintaining their accuracy. These representations must be sufficiently general to ensure compatibility with a broad range of computational tools while also enabling a detailed and rigorous analysis of the underlying problem.
- **Decoupling Interpretation from Processing:** The principle of decoupling problem interpretation from the underlying processing mechanisms ensures that result interpretation is deferred to later stages. This approach allows the computational process to concentrate solely on executing the necessary calculations and optimizations, enhancing efficiency and flexibility.
- **Minimization of Processing Tools Adaptation:** The problem is abstracted and represented in a manner that minimizes the need for extensive adaptation of processing tools. This approach ensures that compatibility with existing computational methods and tools arises naturally, eliminating the necessity for significant modifications.
- **Scalability:** The problem representation must be designed with scalability in mind, ensuring that the approach remains both effective and efficient across varying problem sizes, from small-scale instances to large, complex systems.

## III. Literature Review

Software engineering has been shaped by several innovative paradigms aimed at improving the development process, including Component-Based Software Engineering (CBSE) [5], Model-Driven Engineering (MDE) [6] and Aspect-Oriented Software Development (AOSD) [7]. These paradigms primarily focus on modularity, automation, and development management—each proposing structured methodologies for software design, implementation, and evolution.

However, these paradigms share a common limitation: they are designed for development-centric tasks, meaning they emphasize software construction, modularization, and maintenance strategies rather than focusing on the core computational formulation of software engineering problems. CBSE, for instance, promotes modular component reuse but does not address software optimization as a general computational problem. MDE structures development around models and transformations but remains highly dependent on domain-specific configurations and tools. AOSD isolates cross-cutting concerns but lacks a structured mechanism for problem abstraction beyond aspect separation.

In contrast, ASE is not a development framework—it is a problem-centric paradigm. Rather than prescribing how software should be developed, ASE focuses on how software engineering problems can be optimally represented and solved using mathematical, machine learning, and algorithmic techniques. ASE introduces a fundamental shift by decoupling problem modeling from the processing mechanism [8], [9], ensuring that complex software engineering challenges—such as design pattern detection and refactoring—can be tackled independently of specific tools and heuristic methodologies. A core advantage of ASE is its ability to abstract software artifacts into computationally efficient representations, allowing them to be processed using a wide range of tools without requiring tool-specific adaptations.

### A. Key Differences Between ASE and Development Paradigms

1) **Purpose**:
- ASE is a *problem-centric design paradigm* focused on the *formulation, encoding, and resolution of complex software engineering challenges*. It achieves this by employing *mathematical abstractions, compact representations, and ensuring compatibility with computational tools*.
- Development Paradigms (e.g., MDE, CBSE, AOSD, SBSE): These paradigms are primarily concerned with *guiding the design, development, and maintenance of software systems*. Their focus lies in *establishing structured workflows, processes, and toolchains that facilitate software construction and evolution*.

2) **Scope**:
- ASE: Primarily addresses *specific analytical challenges*, including *code structure optimization, design pattern recognition, and software artifact abstraction*. It aims to *generalize problem representations* to enable *efficient computational processing*.
- Development Paradigms: Encompass *the entire software lifecycle*, covering phases from *requirements engineering to system deployment*, with an emphasis on the *methodologies and frameworks that govern software development and evolution*.

3) **Nature of Application**:
- ASE: Operates as a *computational, problem-oriented approach*, facilitating the *integration of advanced analytical*

*techniques*, such as *search-based refactoring and machine learning*, to *analyze and optimize software artifacts*.
- Development Paradigms: Are inherently *process-centric*, providing *structured guidelines and workflows* that enable software teams to *collaborate and systematically develop software systems*.

4) **Flexibility**:
- ASE: Is *method-agnostic* and can be *seamlessly integrated with existing development paradigms*. For instance, *ASE principles can be applied within an MDE workflow* to enhance problem representation and analysis.
- Development Paradigms: Typically *define a rigid set of processes and rules* for software development, which can, in some cases, *limit adaptability and flexibility*.

5) **ASE's Position in Software Engineering**: While paradigms such as *MDE and CBSE* emphasize *system design and implementation*, ASE complements these approaches by:
- Providing *efficient problem representations* that enhance computational processing.
- Ensuring *compatibility with diverse tools* to facilitate *optimization and analysis*.
- Introducing a *language-agnostic framework* for problem abstraction, making it applicable across different software engineering challenges.

### B. How ASE Complements The Other Paradigms

To illustrate how ASE aligns with and enhances existing paradigms, *Table I* presents a *detailed comparative analysis* between ASE and traditional paradigms such as *MDE, CBSE, and AOSD*, across key dimensions.

### C. ASE as a Pioneering Paradigm

Unlike existing paradigms, which focus on development workflows, ASE is problem-first: it does not dictate how software should be written but rather how software engineering problems should be formulated and solved efficiently. By emphasizing abstraction, generalization, and computational scalability, ASE provides a framework that is tool-agnostic, extensible, and adaptable across a broad range of software engineering challenges.

To the best of our knowledge, ASE represents the first systematic effort to treat software engineering problems as structured computational challenges, bridging the gap between formal problem representation and practical software optimization. This paradigm shift unlocks new possibilities for integrating AI, heuristic optimization, and mathematical modeling into software engineering, setting the stage for a new generation of analytical, tool-independent approaches.

## IV. METHODOLOGY

In this study, the methodology is organized around a set of use cases illustrating the practical application of the ASE paradigm to complex software engineering challenges. Rather than centering on a single procedural method, the ASE paradigm is substantiated by examining the outcomes produced by the ASE-based frameworks, BSS [13] and ODR [14], as well as by applying these frameworks in two distinct domains:
- Design Pattern Detection (DPD) [15], [16]: Demonstrating ASE's ability to encode and analyze software structures.
- Search-Based Software Refactoring (SBR) [17], [18]: Showcasing ASE's role in optimizing software artifacts through heuristic-driven transformations.

By framing the methodology around these use cases, we illustrate ASE's **generality, scalability, and compatibility** with computational tools.

### A. Use Case 1: Design Pattern Detection

*1) Problem Definition and Challenges:* Design pattern detection is essential for software *maintenance and assessment*, but traditional approaches face several challenges, including:
- **Static nature of existing methods** – Lack of behavioral and contextual analysis.
- **Difficulty in evaluating state-of-the-art tools** – Limited cross-method compatibility.
- **Language dependency** – Existing methods do not generalize well across programming languages.
- **Scalability issues** – Inefficiencies in handling large datasets and lack of training data for DPD tasks.

*2) Use Case 1: ASE's Role in Addressing These Challenges:* The ASE is employed in the development of BSS, a compact, language-agnostic encoding designed to address the limitations of existing DPD methodologies. The BSS framework leverages ASE principles to enhance the efficacy and accuracy of pattern detection approaches. A comprehensive overview of how these principles are applied to overcome current DPD challenges is presented in Figure 2.

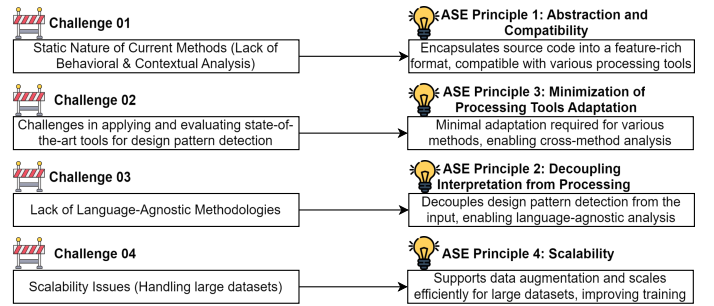

Fig. 2: Visualizing the Resolution of Key Challenges in Design Pattern Detection through ASE Principles.

### B. Use Case 2: Search-Based Software Refactoring

*1) Problem Definition:* Search-based software refactoring aims to optimize software structure by directly refactoring the codebase using heuristics and optimization methods, but existing approaches face several challenges:
- **Limited versatility in problem encoding** – High dependency on tool-specific encodings.

TABLE I: Comparison of ASE with existing software development paradigms.

| Criteria | ASE | MDE [10] | CBSE [11] | AOSD [12] |
|---|---|---|---|---|
| **Problem Abstraction** | High | High | Medium | Low |
| **Tool Compatibility** | High | Medium (platform-specific) | Medium (framework-specific) | Low (aspect weavers only) |
| **Scalability** | High | Medium (limited by model size) | High | High |
| **DPD support** | ✓ | Medium (model-based analysis) | Low | Low |
| **Refactoring Support** | ✓ | Medium (model-based refactoring) | ✗ | ✗ |
| **Behavioral Analysis** | ✓ | ✗ | Low | ✗ |
| **Language Agnosticism** | ✓ | ✓ | Medium | ✗ |
| **Data Augmentation Support** | ✓ | Low | Low | Low |

- **Difficulty in evaluating state-of-the-art heuristics** – Poor compatibility across multiple search techniques.
- **High computational overhead** – Iterative transformations increase processing time.
- **Scalability issues** – Inefficiencies in handling large-scale software systems.

*2) ASE's Role in Addressing These Challenges:* The ASE is utilized in the design and development of the ODR framework, which integrates artifact and solution representations to minimize computational overhead and enhance scalability. A detailed illustration of how ASE principles address existing SBR challenges is provided in Figure 3.

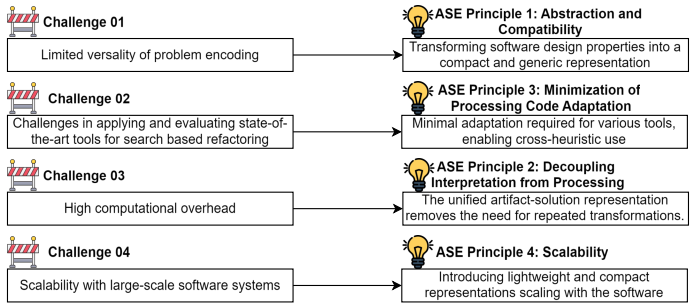

Fig. 3: Visualizing the Resolution of Key Challenges in Search-Based Refactoring through ASE Principles.

## V. EMPIRICAL EVALUATION AND DISCUSSION OF THE ASE PARADIGM APPROACHES

This section conducts an empirical evaluation of ASE by examining its performance in two distinct software engineering contexts: DPD and SBR. To validate ASE's feasibility, effectiveness, generality, and efficiency, we present a series of quantitative experiments involving the ODR and BSS frameworks across both DPD and SBR scenarios.

### A. Experimental Results

*1) Results for Design Pattern Detection using BSS:* Table II presents the empirical evaluation of various design pattern detection methods, comparing the ASE-based *BSS* approach with existing methods, including *DPD_Att*, *DPD_F*, and *FeatureMap*. The results highlight the significant advantages introduced by ASE principles in enhancing both accuracy and generalizability.

*a) Performance Comparison:* The BSS approach achieves an accuracy of 92.52%, significantly outperforming prior methods. Among the evaluated approaches, BSS (ASE-based) achieves the highest accuracy at 92.52%, followed by DPD_Att with 86%, which employs a transformer-based model but lacks behavioral analysis. The DPD_F method attains an F1-score of 80.75%, relying on feature-based detection but lacking structured encoding, while FeatureMap records the lowest F1-score of 52.93%, focusing on feature visualization rather than structured detection. Beyond accuracy, the BSS model also demonstrates superior precision (92.55%) and recall (92.52%), ensuring a balanced and robust detection capability across different design patterns. In contrast, *DPD_F* and *FeatureMap* exhibit significantly lower recall, suggesting a higher false-negative rate in design pattern identification, further highlighting the effectiveness of the BSS approach.

*b) Contribution of ASE Principles:* The superior performance of BSS can be attributed to the following ASE principles, which enhance the methodology by addressing key challenges in design pattern detection. One of the major limitations of existing methods (*DPD_Att, DPD_F, FeatureMap*) is their reliance solely on structural or statistical features, failing to capture crucial behavioral aspects of design patterns. ASE's Abstraction and Compatibility principle enables *BSS* to incorporate both behavioral and structural representations, making it the only method in the comparison to provide this capability. Another notable shortcoming of prior methods is the absence of comprehensive training datasets, which restricts their capacity to generalize across multiple programming languages. By utilizing ASE's Scalability and Decoupling principles, *BSS* incorporates language-agnostic data augmentation techniques, thereby enhancing model robustness. This advantage is evident in the larger dataset size adopted by BSS relative to (DPD_Att) and (DPD_F, FeatureMap).

Additionally, traditional methods suffer from language de-pendency, restricting their usability to specific programming languages. ASE's Decoupling Interpretation from Processing principle ensures that the *BSS* model remains language-agnostic, enabling seamless design pattern detection across programming languages, whereas other methods remain tied to a single language. Furthermore, previous methods only detect a subset of design patterns—*DPD_Att* and *DPD_F* cover 13 patterns, while *FeatureMap* detects only 8. In contrast, BSS achieves full coverage of all 23 GoF design patterns, ensuring a more comprehensive and reliable detection framework. These results validate ASE's structured problem formulation as a transformative approach to design pattern detection, outperforming traditional methodologies in terms of accuracy, generalizability, and scalability.

*2) Results for Search-Based Refactoring using ODR:*

*a) Comparison of Refactoring Methods:* The empirical evaluation of SBR methods, presented in Table III, compares the ODR framework against AST-Based Approaches (ATBA) and Graph-Based Approaches (GBA). The results highlight ODR's efficiency, scalability, and adaptability, demonstrating its ability to optimize software artifacts while minimizing computational overhead through ASE principles.

One of the key advantages of ODR is its ability to eliminate the need for separate solution representation, significantly reducing computational complexity compared to traditional refactoring methods. Additionally, it exhibits high genericity, making it more adaptable across diverse optimization techniques, whereas ATBA and GBA approaches tend to be restricted by their encoding-specific dependencies. The ODR framework also introduces lower memory usage, allowing it to scale effectively for large projects, a crucial factor in handling real-world software repositories. Furthermore, it provides comprehensive tracking of object associations, inheritance, and interface implementations, ensuring a more structured and coherent refactoring process that surpasses both ATBA and GBA in restructuring efficiency.

*b) Contribution of ASE Principles:* The effectiveness of ODR stems from its foundation in ASE principles, which address key challenges in existing refactoring approaches by enhancing computational efficiency, adaptability, and scalability. One major limitation in AST-Based Approaches (ATBA) and Graph-Based Approaches (GBA) is the need for separate solution representations, which significantly increase memory consumption and computational complexity. ASE's Decoupling Interpretation from Processing principle enables ODR to eliminate explicit solution representations, thereby reducing computational overhead and improving the overall efficiency of the refactoring process.

Another major drawback of traditional approaches is their lack of genericity, as both ATBA and GBA models depend on encoding-specific representations, restricting their adaptability to various heuristic techniques. By leveraging ASE's Abstraction and Compatibility principle, ODR achieves high genericity, making it applicable across diverse codebases without requiring extensive modifications. Moreover, traditional refactoring techniques tend to be memory-intensive, requiring large-scale in-memory structures that lead to increased computational resource consumption. ODR, guided by ASE's Abstraction and Scalability principles, optimizes artifact encoding to ensure low memory usage, thereby maintaining scalability while preserving refactoring effectiveness.

Despite these improvements, one of the remaining challenges of ODR is its limited explainability and code behavior preservation, particularly when compared to some graph-based approaches that offer more explicit semantic tracking. ASE's Scalability principle addresses this limitation by prioritizing structural integrity in refactoring while minimizing excessive transformation complexity. However, future research should focus on improving the interpretability of refactored solutions and integrating behavior-preserving techniques to enhance trustworthiness in software transformation. Overall, these findings validate ASE's ability to significantly advance refactoring methodologies, offering a structured, scalable, and computationally efficient alternative to conventional approaches.

## VI. LIMITATIONS AND FUTURE DIRECTIONS

While the ASE design paradigm has demonstrated significant improvements in addressing complex software engineering challenges, certain limitations remain that warrant further exploration.

### A. Explainability of Abstraction and Manipulation

One notable limitation of the ASE paradigm is the reduced explainability inherent in its compact abstraction format and manipulation process, which can obscure how specific decisions influence final outcomes, potentially reducing trust in the paradigm. To address this, Explainable Artificial Intelligence (XAI) techniques can be integrated into ASE to improve interpretability. Approaches such as attention visualization can highlight influential components in abstraction by mapping attention weights to original code or design patterns, while feature importance scoring and saliency maps can reveal which encoded features drive system predictions. Additionally, interactive explanation dashboards can provide users with real-time insights into how changes in parameters affect the abstraction process. Model-agnostic methods like LIME [30] (Local Interpretable Model-agnostic Explanations) and SHAP [31] (SHapley Additive exPlanations) can further enhance transparency by offering localized insights into individual decisions. By incorporating these XAI features, ASE can improve its explainability without sacrificing computational efficiency, fostering greater trust, facilitating debugging, and refining the abstraction process to better align abstract representations with practical software engineering applications.

### B. Scalability in Data-Driven Environments

The proliferation of data-driven systems, especially those handling petabyte-scale data, presents additional challenges. Although ASE is designed for computational efficiency, processing and analyzing extremely large datasets introduces issues related to performance optimization, resource management, and scalability. Future research should investigate how

TABLE II: A qualitative and quantitative evaluation of the BSS framework is presented in comparison with three leading approaches. This analysis assesses key performance metrics such as accuracy, F1-score, and design pattern coverage, while also examining methodological distinctions and practical benefits.

| Metrics | BSS (ASE-Based) [13] | DPD_Att[19] | DPD_F [20] | FeatureMap [21] |
|---|---|---|---|---|
| **Accuracy** (%) | **92.52** | 86 | – | – |
| **F1 Score** (%) | **92.47** | 86 | 80.75 | 52.93 |
| **Precision** (%) | **92.55** | 87 | 81.44 | 52.30 |
| **Recall** (%) | **92.52** | 86 | 80.47 | 54.42 |
| **Dataset** | PyDesignNet | DPD-F-Corpus++ | DPD-F-Corpus | DPD-F-Corpus |
| **GoF Design Pattern Coverage** | 23 | 13 | 13 | 8 |
| **Dataset size** | 1832 files | 1645 files | 1300 files | 1300 files |
| **Behavioral Analysis** | ✓ | ✗ | ✗ | ✗ |
| **Data augmentation possibility** | ✓ | ✗ | ✗ | ✗ |
| **Language agnostic** | ✓ | ✗ | ✗ | ✗ |

TABLE III: Comparison of qualitative attributes among ODR, ATBA, and GBA approaches.

| Attributes | ODR (ASE-Based) [14] | ATBA [22][23][24][25] | GBA[26][27][28][29] |
|---|---|---|---|
| Solution Representation | Not required | Often Required | Often Required |
| Genericity[2] | High | Limited | Limited |
| Memory Usage | Low | High | Moderate |
| Objects Associations Encoding | ✓ | ✓ | ✓ |
| Encapsulation States Encoding | ✗ | ✗ | ✗ |
| Inheritance Tracking | ✓ | ✓ | Dependent on Context |
| Interface Implementing Tracking | ✓ | ✓ | Dependent on Context |
| Method Decorators Tracking | ✗ | ✓ | ✗ |
| Method Signature Analysis | ✓ | ✓ | ✓ |
| Code Semantics and Syntax Analysis | ✗ | Dependent on Context | ✗ |
| Explainability of Refactorings | Limited | Limited | Moderate |
| Refactoring Ability | High | Limited | Moderate |
| Code Behavior Preservation[3] | Limited | Dependent on context | Limited |

ASE can be integrated with distributed computing techniques, data reduction strategies, and other scalable solutions to maintain its effectiveness in environments with vast amounts of data.

## VII. CONCLUSION

This paper introduces and evaluates the ASE paradigm, which provides a *problem-centric approach* to modeling and solving complex software engineering challenges. Unlike traditional development paradigms that primarily focus on software construction and lifecycle management, ASE emphasizes structured problem formulation, abstraction, and computational compatibility, enabling efficient integration with a wide range of analytical techniques and optimization methods.

The empirical evaluation of ASE demonstrates its impact in two distinct use cases: Design Pattern Detection (DPD) and Search-Based Refactoring (SBR). In DPD, ASE facilitates the development of BSS, a language-agnostic encoding that captures both *structural and behavioral* features, achieving a 6.5% increase in accuracy compared to prior approaches. In SBR, ASE enables the ODR framework, which reduces computational overhead, improves refactoring efficiency, and enhances structural integrity, leading to a reduction in optimization time while improving maintainability metrics such as CBO and LCOM.

While ASE offers significant improvements in accuracy, scalability, and cross-tool compatibility, several challenges remain. The paradigm currently exhibits limited explainability due to its reliance on compact encoded representations, raising the need for XAI techniques [32] integration. Additionally, behavioral tracking in refactoring remains an open challenge, requiring future research into behavior-preserving optimization techniques. Furthermore, computational complexity trade-offs, particularly in large-scale software projects, must be addressed through adaptive feature reduction strategies.

Looking ahead, ASE's structured encoding principles hold promise for broader applications, including automated code quality assessment, evolutionary software optimization [33], and software verification [34]. By integrating emerging AI-driven interpretability techniques, ASE can evolve into a more transparent and adaptive framework, paving the way for future innovations in software engineering analytics.

The conceptual underpinnings and findings outlined in this paper position ASE as a versatile and computationally efficient paradigm, bridging the gap between abstract software engineering concepts and practical, tool-compatible optimization approaches. This paper provides a foundation for future work focused on refining ASE's core principles and extending its applicability to emerging areas in analytical software engineering.

## REFERENCES

[1] Smite, D., Wohlin, C., Gorschek, T. et al. Empirical evidence in global software engineering: a systematic review. *Empir Software Eng 15, 91–118 (2010)*. https://doi.org/10.1007/s10664-009-9123-y

[2] P. Devanbu, T. Zimmermann, and C. Bird, "Belief & Evidence in Empirical Software Engineering," in *Proc. 38th Int. Conf. on Software Engineering*, New York, NY, USA, May 2016, pp. 108–119.

[3] M. Saari, M. Nurminen and P. Rantanen, "Survey of Component-Based Software Engineering within IoT Development,"*45th Jubilee International Convention on Information, Communication and Electronic Technology (MIPRO), Opatija, Croatia, 2022, pp. 824-828*, doi: 10.23919/MIPRO55190.2022.9803785.

[4] L. Burgueño, A. Burdusel, S. Gérard and M. Wimmer, "Preface to MDE Intelligence 2019: 1st Workshop on Artificial Intelligence and Model-Driven Engineering," *ACM/IEEE 22nd International Conference on Model Driven Engineering Languages and Systems Companion (MODELS-C), Munich, Germany, 2019, pp. 168-169*, doi: 10.1109/MODELS-C.2019.00028.

[5] T. Vale *et al.*, "Twenty-Eight Years of Component-Based Software Engineering," *J. Syst. Softw.*, vol. 111, pp. 128–148, Jan. 2016.

[6] J. Whittle, J. Hutchinson, and M. Rouncefield, "The State of Practice in Model-Driven Engineering," *IEEE Softw.*, vol. 31, no. 3, pp. 79–85, May 2014.

[7] A. Rashid, P. Matos, R. Chitchyan, and B. Farbey, "Aspect-Oriented Software Development in Practice: Tales from AOSD-Europe," *Computer*, vol. 43, no. 2, pp. 19–26, Feb. 2010.

[8] D. R. Hofstadter, *Gödel, Escher, Bach: An Eternal Golden Braid*. Reading, MA, USA: Addison-Wesley, 1999.

[9] W. J. Rapaport, *Philosophy of Computer Science*. Hoboken, NJ, USA: Wiley-Blackwell, 2023.

[10] J. Cabot, R. Clariso, E. Guerra, and M. Wimmer, "Cognifying Model-Driven Software Engineering," in *Proc. Software Technologies: Applications and Foundations*, 2018, pp. 154–160.

[11] I. Crnkovic, "Component-Based Software Engineering — New Challenges in Software Development," *Software Focus*, vol. 2, no. 4, pp. 127–133, 2001.

[12] M. Khaari and R. Ramsin, "Process Patterns for Aspect-Oriented Software Development," in *Proc. 17th IEEE Int. Conf. on Engineering of Computer Based Systems*, Mar. 2010, pp. 241–250.

[13] T. Houichime and Y. El Amrani, "Context is All You Need: A Hybrid Attention-Based Method for Detecting Code Design Patterns," *IEEE Access*, pp. 101–120, 2025.

[14] T. Houichime and Y. El Amrani, "Optimized Design Refactoring (ODR): A Generic Framework for Automated Search-Based Refactoring to Optimize Object-Oriented Software Architectures," *Autom. Softw. Eng.*, vol. 31, no. 2, p. 48, Jun. 2024.

[15] J. Asaad and E. Avksentieva, "A Review of Approaches to Detecting Software Design Patterns," in *Proc. 35th Conf. of Open Innovations Association (FRUCT)*, Tampere, Finland, Apr. 2024, pp. 142–148.

[16] B. B. Mayvan, A. Rasoolzadegan, and Z. Ghavidel Yazdi, "The state of the art on design patterns: A systematic mapping of the literature," *J. Syst. Softw.*, vol. 125, pp. 93–118, Mar. 2017.

[17] T. Mariani and S. R. Vergilio, "A Systematic Review on Search-Based Refactoring," *Inf. Softw. Technol.*, vol. 83, pp. 14–34, Mar. 2017.

[18] A. A. B. Baqais and M. Alshayeb, "Automatic Software Refactoring: A Systematic Literature Review," *Softw. Qual. J.*, vol. 28, no. 2, pp. 459–502, Jun. 2020.

[19] R. Mzid, I. Rezgui, and T. Ziadi, "Attention-based method for design pattern detection," in *Software Architecture*, M. Galster *et al.*, Eds. Cham, Switzerland: Springer, 2024, pp. 86–101.

[20] N. Nazar, A. Aleti, and Y. Zheng, "Feature-based software design pattern detection," *J. Syst. Softw.*, vol. 185, p. 111179, Mar. 2022.

[21] H. Thaller, L. Linsbauer, and A. Egyed, "Feature Maps: A Comprehensible Software Representation for Design Pattern Detection," in *Proc. 26th IEEE Int. Conf. on Software Analysis, Evolution and Reengineering (SANER)*, Feb. 2019, pp. 207–217.

[22] S. Ghaith and M. Ó Cinnéide, "Improving Software Security Using Search-Based Refactoring," in *Proc. Int. Symp. on Search-Based Softw. Eng.*, vol. 7515, Sep. 2012, p. 135.

[23] I. H. Moghadam and M. Ó Cinnéide, "Automated Refactoring Using Design Differencing," in *Proc. Euromicro Conf. on Software Maintenance and Reengineering (CSMR)*, Mar. 2012, p. 52.

[24] M. O'Keeffe and M. Ó Cinnéide, "Search-Based Software Maintenance," *J. Syst. Softw.*, vol. 81, p. 260, Apr. 2006.

[25] M. O'Keeffe and M. Ó Cinnéide, "Getting the Most from Search-Based Refactoring," in *Proc. 9th Annu. Conf. on Genetic and Evolutionary Computation (GECCO)*, Jul. 2007, p. 1120.

[26] D. Di Pompeo and M. Tucci, "Multi-Objective Software Architecture Refactoring Driven by Quality Attributes," *arXiv*, Jan. 2023.

[27] S. Herold and M. Mair, "Recommending Refactorings to Re-establish Architectural Consistency," in *Software Architecture*, P. Avgeriou and U. Zdun, Eds. Cham, Switzerland: Springer, 2014, pp. 390–397.

[28] S. Kebir, I. Borne, and D. Meslati, "A Genetic Algorithm-Based Approach for Automated Refactoring of Component-Based Software," *Inf. Softw. Technol.*, vol. 88, pp. 17–36, Aug. 2017.

[29] H. Masoud and S. Jalili, "A Clustering-Based Model for Class Responsibility Assignment Problem in Object-Oriented Analysis," *J. Syst. Softw.*, vol. 93, pp. 110–131, Jul. 2014.

[30] M. J. Hjuler, L. H. Clemmensen and S. Das, "Exploring Local Interpretable Model-Agnostic Explanations for Speech Emotion Recognition with Distribution-Shift," ICASSP 2025 - *IEEE International Conference on Acoustics, Speech and Signal Processing (ICASSP), Hyderabad, India, 2025, pp. 1-5*, doi: 10.1109/ICASSP49660.2025.10889825.

[31] S. M. Gbashi, O. O. Olatunji, P. A. Adedeji and N. Madushele, "An Explainable AI Approach Using SHapley Additive exPlanations for Feature Selection in Vibration-Based Fault Diagnostics of Wind Turbine Gearbox," *IEEE 5th International Conference on Electro-Computing Technologies for Humanity (NIGERCON), Ado Ekiti, Nigeria, 2024, pp. 1-5*, doi: 10.1109/NIGERCON62786.2024.10927384.

[32] David Gunning et al. ,XAI—Explainable artificial *intelligence.Sci. Robot.4*,eaay7120(2019).DOI:10.1126/scirobotics.aay7120

[33] Timo Mantere, Jarmo T. Alander, Evolutionary software engineering, a review, Applied Soft Computing, *Journal of Systems and Software*,Volume 5, Issue 3,2005, Pages 315-331,.DOI:10.1016/j.asoc.2004.08.004

[34] N. Rajabli, F. Flammini, R. Nardone and V. Vittorini, "Software Verification and Validation of Safe Autonomous Cars: A Systematic Literature Review," in IEEE Access, vol. 9, pp. 4797-4819, 2021, doi: 10.1109/ACCESS.2020.3048047.